\documentclass[a4paper,12pt]{article}

\usepackage{graphicx}
\usepackage{amsmath}
\usepackage{bm,cite}

\begin{document}

\begin{center}
{\large\bf DIS at small $\bm{x}$ and hadron-hadron scattering at high energies via the holographic Pomeron exchange}

{Akira Watanabe$^{1,2,\P}$}

$^1${Institute of High Energy Physics and Theoretical Physics Center for Science Facilities, Chinese Academy of Sciences, Shijingshan, 100049, Beijing, People's Republic of China}

$^2${University of Chinese Academy of Sciences, Shijingshan, 100049, Beijing, People's Republic of China}

$^\P${E-mail: akira@ihep.ac.cn}
\end{center}

\centerline{\bf Abstract}
We present our analysis on the high energy scattering processes in the framework of the holographic QCD, assuming that the Pomeron exchange gives a dominant contribution to the cross sections.
Focusing on the two-body scattering, we employ the Brower-Polchinski-Strassler-Tan Pomeron exchange kernel to describe the strong gluonic interaction, and utilize the bottom-up AdS/QCD models to obtain the density distributions of the involved hadrons in the five-dimensional AdS space, to calculate the total cross sections.
Considering the unpolarized deep inelastic scattering at small Bjorken $x$ and the high energy hadron-hadron scattering, we show that our calculations are consistent with the presently available experimental data.
Our predictions made via this framework can be tested at various experimental facilities in the future. 

\section{Introduction}
Experiments of high energy scattering phenomena have provided us valuable opportunities to investigate the partonic structure of hadrons, which is one of the most important subjects in high energy physics.
The fundamental theory of the strong interaction among the quarks and the gluons is known as quantum chromodynamics (QCD), and this theory was established several decades ago.
However, even now it is extremely hard to study the quark-gluon dynamics in the nonperturbative kinematic region by directly using QCD, and the analysis by effective approaches are necessary in such a regime.

The total cross sections are basically observable in high energy scattering experiments, but the parton distribution functions (PDFs) of the involved hadrons are in principle required as inputs to theoretically predict those.
PDFs are not calculable because of the nonperturbative nature, but it is still possible to directly calculate total cross sections by utilizing effective models.
Hence, via model studies confronted with the experimental data, we may obtain a better understanding of PDFs.

In this report, based on Refs.~\cite{Watanabe:2012uc,Watanabe:2013spa,Watanabe:2015mia,Watanabe:2018owy} we present our analysis on total cross sections of the two two-body scattering processes, the unpolarized deep inelastic scattering (DIS) at small Bjorken $x$ and the high energy hadron-hadron scattering, assuming the Pomeron exchange to describe the dominant gluonic interaction in the framework of holographic QCD.
We employ the Brower-Polchinski-Strassler-Tan (BPST) Pomeron exchange kernel~\cite{Brower:2006ea}, which gives a Pomeron exchange contribution to the cross section, to realize this.
The probe photon in DIS is described by the wave function of the U(1) vector field~\cite{Polchinski:2002jw}, and the density distributions of the involved hadrons are obtained by utilizing the bottom-up AdS/QCD models~\cite{Erlich:2005qh,Hong:2006ta} in the five dimensional AdS space.

After the model setup is briefly explained, firstly we present our results for the electron-nucleon DIS at small $x$.
It is shown that the experimental data measured at HERA can be well reproduced within our model.
Since a few adjustable model parameters are determined with the proton $F_2$ structure function data, the longitudinal structure function $F_L$ can be predicted without any additional parameter.
The most important advantage of our framework is that this method is applicable to analysis on structure functions of other hadrons, giving the appropriate distribution functions of the hadrons in the AdS space.
As examples, besides the resulting pion structure function, the results for the photon structure function are also presented.
In the electron-photon DIS, the hadronic component becomes dominant in the small $x$ region.

Secondly, we present another application of the model for the analysis on the total cross section of the nucleon-nucleon collision at high energies.
Although the kinematics is different from that of DIS, our calculation is in agreement with the experimental data, including the ones recently measured by the TOTEM collaboration at LHC.
Similar to the DIS case, the pion involved processes, for examples, are also considered.
Through all these applications, one may see the broad applicability of the present model.

\section{Holographic description of total cross sections}
By virtue of the optical theorem, the total cross section can be calculated from the scattering amplitude in the forward limit:
\begin{equation}
\sigma _{\rm tot} ( s ) = \frac{1}{s} \hspace{1mm} {\rm Im} \hspace{0.3mm} \mathcal{A} (s, t = 0) , \label{eq:ot}
\end{equation}
where $s$ and $t$ are the Mandelstam variables.
Considering the two-body scattering process, $1 + 2 \to 3 + 4$, and employing the BPST Pomeron exchange kernel $\chi$, the scattering amplitude is expressed in the five-dimensional AdS space as
\begin{equation}
\mathcal{A} (s, t) = 2 i s \int d^2 b \hspace{0.3mm} e^{i \bm{k_\perp } \cdot \bm{b} } \int dz dz' P_{13} (z) P_{24} (z') \left[ 1 - e^{i \chi (s, \bm{b}, z, z')} \right] , \label{eq:sa}
\end{equation}
where $z$ and $z'$ are fifth coordinates, $\bm{b}$ denotes the two-dimensional impact parameter, and $P_{13} (z)$ and $P_{24} (z')$ are density distributions of the involved hadrons.
If the hadron is a normalizable mode, its density distribution is normalized.

Applying Eq.\eqref{eq:ot} and picking up the leading contribution from the eikonal representation in Eq.\eqref{eq:sa}, the total cross section can be written as
\begin{equation}
\sigma_{\rm tot} (s) = 2 \int d^2 b \int dz dz' P_{13} (z) P_{24} (z') {\rm Im} \chi (s, \bm{b}, z, z'). \label{eq:tcs-1}
\end{equation}
In the conformal limit, the analytic form of ${\rm Im} \chi$ can be obtained, and the impact parameter integration in Eq.\eqref{eq:tcs-1} can be analytically performed.
Hence, Eq.\eqref{eq:tcs-1} is rewritten as
\begin{align}
&\sigma_{\rm tot} (s) = \frac{g_0^2 \rho^{3/2} }{8 \sqrt{\pi} } \int dz dz' P_{13} (z) P_{24} (z' ) (z z' ) {\rm Im} [\chi_{\rm c} (s, z, z' )] ,  \label{eq:tcs-2} \\
&{\rm Im} [\chi_{\rm c} (s, z, z' ) ] \equiv e^{(1 - \rho) \tau } e^{ - [ ({\log ^2 z / z'}) / {\rho \tau } ] } / {\tau ^{1/2} } , \label{eq:conf_k}
\end{align}
where ${\tau} = {\log} (\rho z z' s / 2)$, and $\rho$ and $g_0^2$ are adjustable parameters that control the energy dependence and the magnitude of the total cross section, respectively.

In this study, we also consider the modified kernel, in which an added term mimics the confinement effect in QCD, given by
\begin{align}
&{\rm Im} [\chi_{\rm mod} (s, z, z' )] \equiv {\rm Im} [\chi_{\rm c} (s, z, z' ) ] + \mathcal{F} (s, z, z' ) {\rm Im} [\chi_{\rm c} (s, z, z_0 z_0' / z' ) ],  \label{eq:mod_k} \\
&\mathcal{F} (s, z, z' ) = 1 - 2 \sqrt{\rho \pi \tau } e^{\eta^2 } {\rm erfc} (\eta ),
\end{align}
where $\eta = \left[ -\log (z z' / z_0 z_0') + \rho \tau \right] / {\sqrt{\rho \tau }}$, and $z_0$ and $z'_0$ are the sharp cutoffs of the fifth coordinates in the infrared (large $z(z')$) region, which characterize the QCD scale.
A smooth cutoff can also be introduced, and in the case $z_0 z_0'$ in Eq.\eqref{eq:mod_k} is replaced with $z_0^2$.

In the case of DIS, some kinematic factors are taken into account in Eq.\eqref{eq:tcs-2}, and the structure functions are expressed as
\begin{equation}
F_i (x, Q^2) = \frac{g_0^2 \rho^{3 / 2} Q^2}{32 \pi ^{5 / 2}} \int dz dz' P_{13}^{(i)} (z, Q^2) P_{24} (z') (z z') {\rm Im} [\chi_{\rm c}(s, z, z')] , \label{eq:sf}
\end{equation}
where $i = 2, L$, and $Q^2$ represents the four-momentum squared of the probe photon.
The photon density distribution has the $Q^2$ dependence in this case, and $P_{13}^{(2)}$ and $P_{13}^{(L)}$ are for the $F_2$ and $F_L$ structure functions, respectively.

To perform the numerical evaluations, we need to specify the density distributions of the involved photon and hadrons.
For the photon we utilize the wave function of the five-dimensional U(1) vector field~\cite{Polchinski:2002jw}, and for the hadrons the density distributions can be calculated from the bottom-up AdS/QCD models.
As to the nucleon and the pion for examples, since the authors in Refs.~\cite{Abidin:2008hn,Abidin:2009hr} obtained the analytic expressions for the gravitational form factors, we can adopt their results in this study.

\section{Numerical results}
\begin{figure}[tb]
\centering
\includegraphics[width=0.75\textwidth]{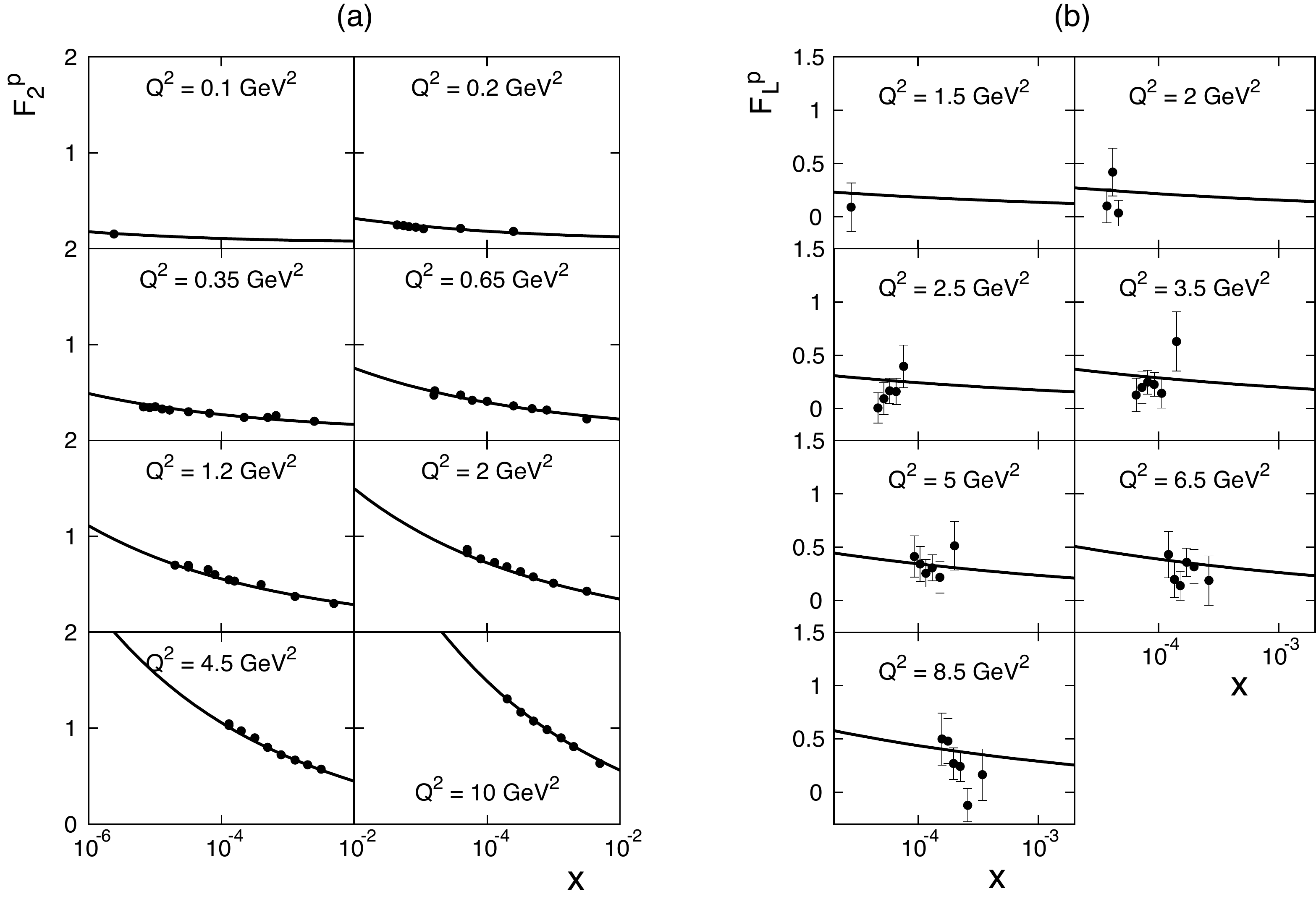}
\caption{
The structure functions, (a) $F_2^p (x, Q^2)$ and (b) $F_L^p (x, Q^2)$, as a function of the Bjorken $x$ for various $Q^2$.
The solid lines represent our results obtained with the modified BPST kernel and the nucleon density distribution calculated by the soft-wall AdS/QCD model.
The experimental data measured at HERA are denoted by the circles with error bars.
}
\label{fig:nucleon_sf}
\end{figure}

\begin{figure}[tb]
\centering
\includegraphics[width=0.53\textwidth]{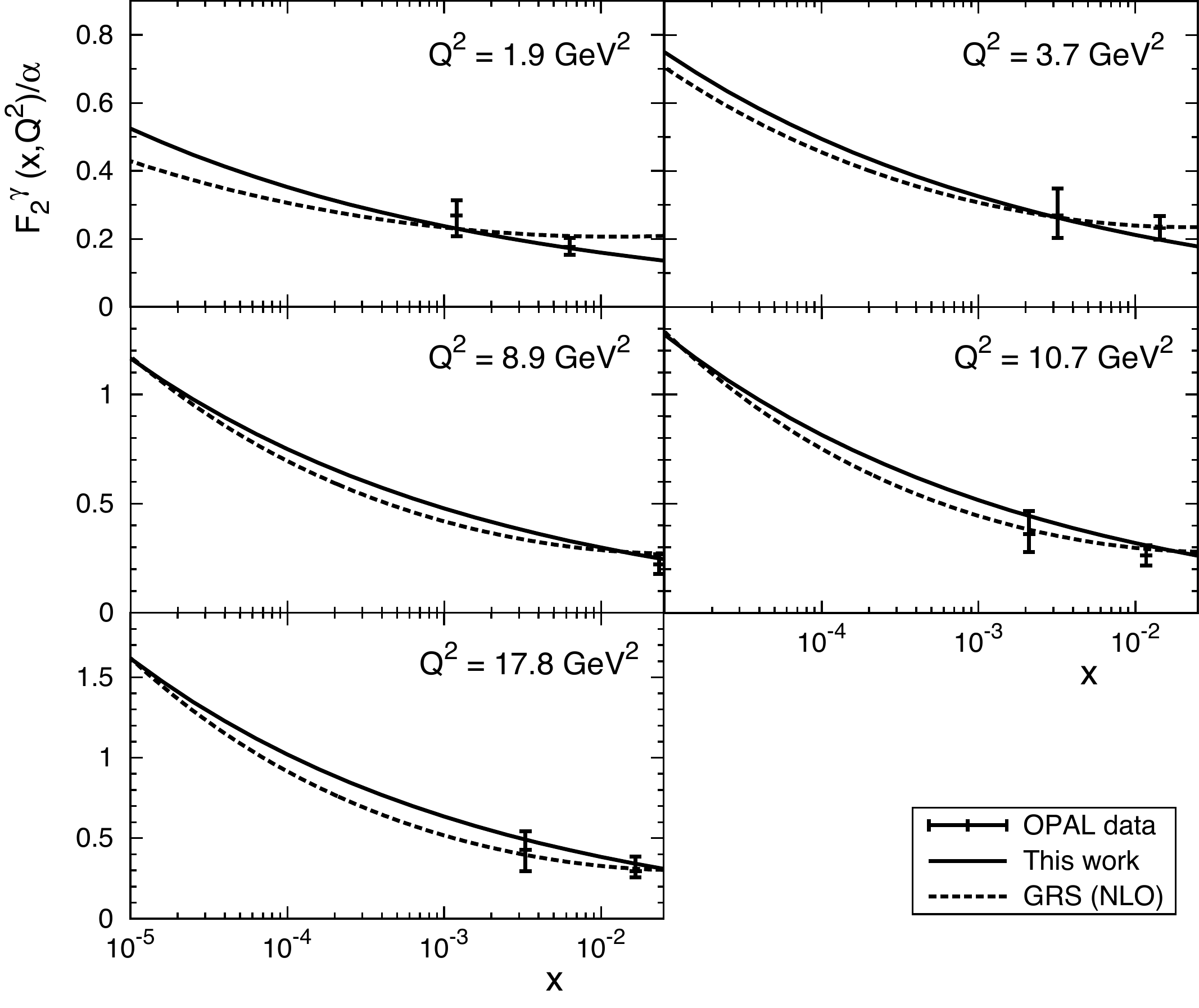}
\caption{
The photon structure function as a function of the Bjorken $x$ for various $Q^2$.
The solid and dashed lines represent our calculations and those obtained from the PDF parameterization~\cite{Gluck:1999ub}, respectively.
The OPAL data are depicted with error bars.
}
\label{fig:photon_f2}
\end{figure}

Here we present our selected results obtained with the modified kernel for the DIS structure functions at small $x$ and the hadron-hadron total cross sections at high energies.
Firstly, we display in Fig.\ref{fig:nucleon_sf}
the resulting nucleon structure functions, compared with the experimental data measured at HERA.
As explained in the first section, the longitudinal structure function can be calculated without any adjustable parameter within the model, and it can be seen from the figure that our calculations are consistent with the data.
The present model setup is applicable to the analysis on the electron-photon DIS, replacing the density distribution of the target.
Our results for the photon $F_2$ are shown in Fig.\ref{fig:photon_f2},
compared with the experimental data measured by the OPAL collaboration at LEP and those from the GRS PDF set~\cite{Gluck:1999ub} at next-to-leading-order (NLO) accuracy.
One can find that our calculations are in agreement with the data and also consistent with those from the PDF parameterization.

\begin{figure}[tb!]
\centering
\includegraphics[width=0.55\textwidth]{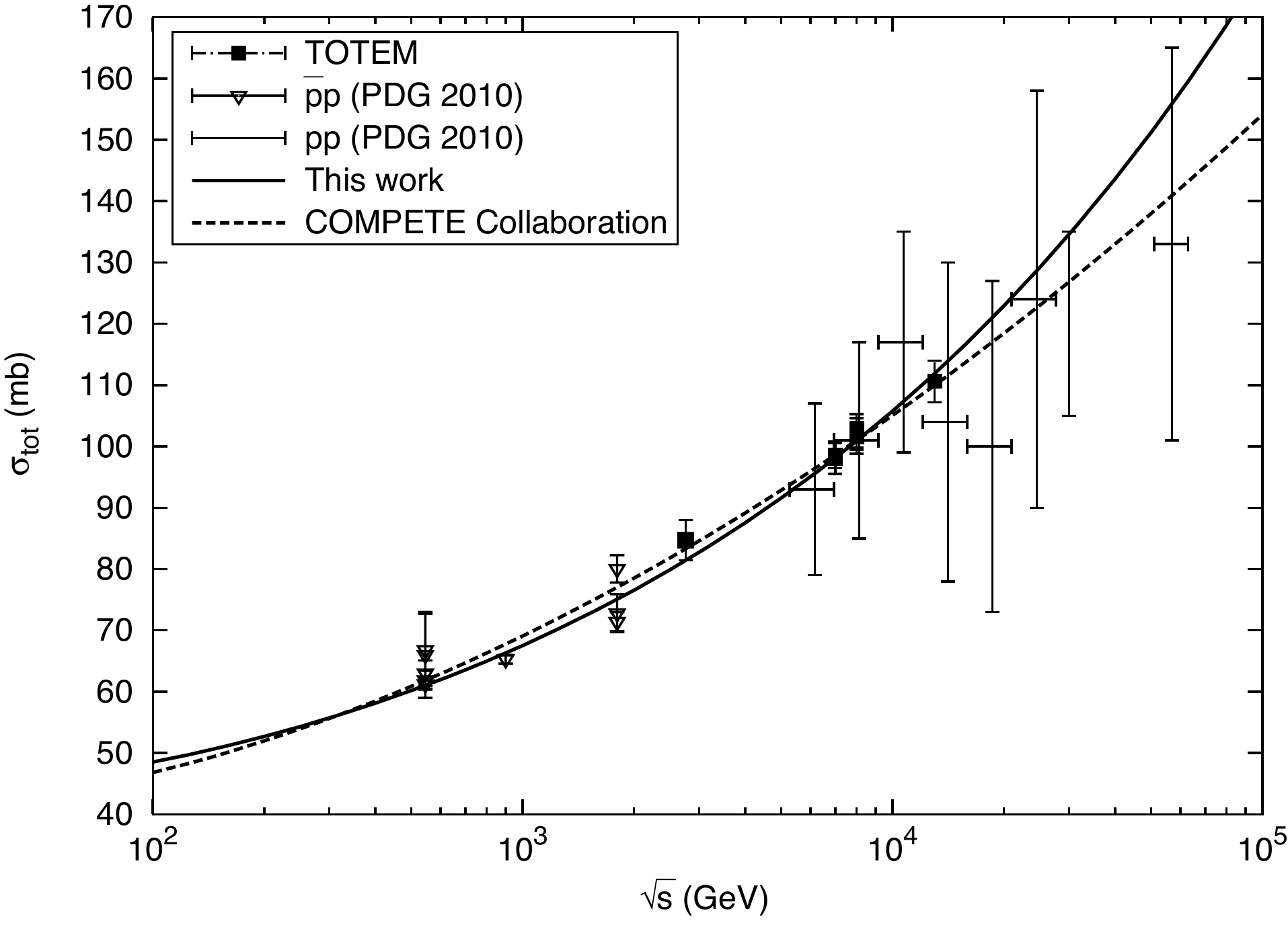}
\caption{
The nucleon-nucleon total cross section as a function of $\sqrt{s}$.
Our calculation is denoted by the solid line, and the dashed line represents the empirical fit by the COMPETE collaboration~\cite{Cudell:2002xe}.
The experimental data are depicted with error bars.
}
\label{fig:nn_tcs}
\end{figure}

Next, we display our result for the nucleon-nucleon total cross section in Fig.\ref{fig:nn_tcs}.
It is seen from the comparison in the figure that our calculation agrees with the data, and consistent with the empirical fit obtained by the COMPETE collaboration~\cite{Cudell:2002xe}, although there is a substantial deviation between the two curves at $\sqrt{s} > 10$~TeV.

\begin{figure}[t!]
\centering
\includegraphics[width=0.73\textwidth]{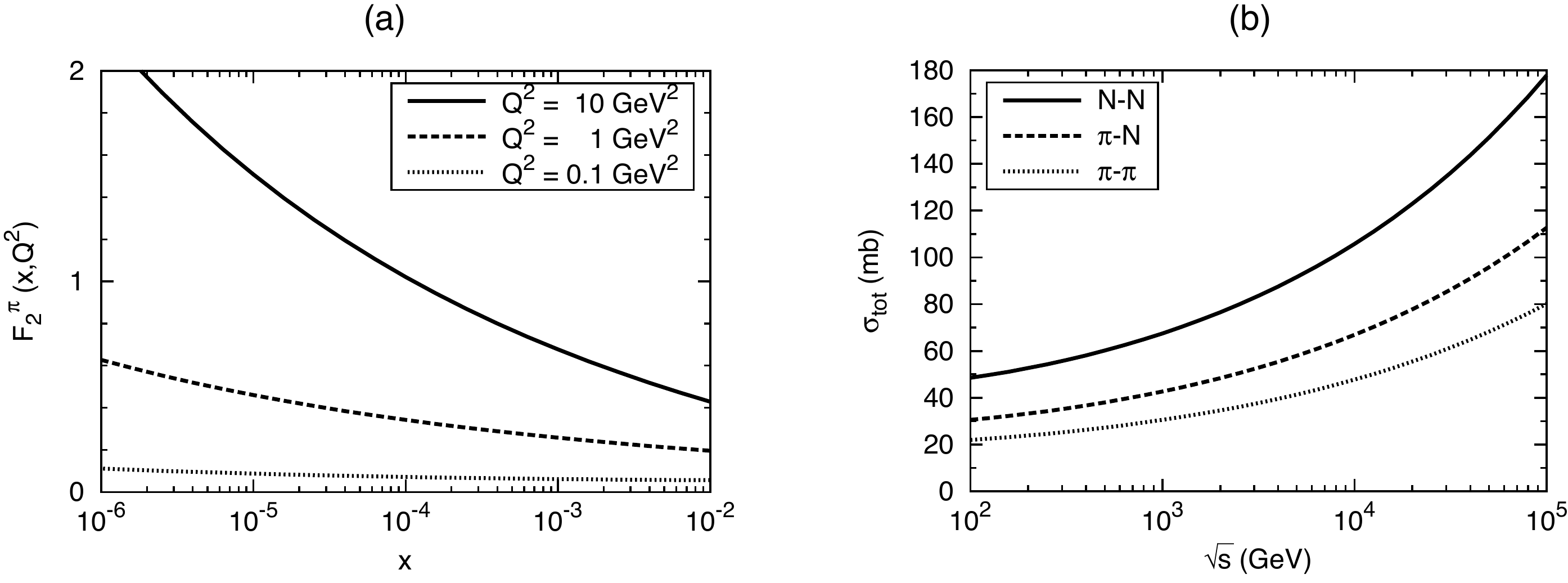}
\caption{
(a) The pion structure function as a function of the Bjorken $x$ for various $Q^2$.
(b) The nucleon-nucleon, pion-nucleon, and pion-pion total cross sections as a function of $\sqrt{s}$.
}
\label{fig:pion_results}
\end{figure}

Finally, we show in Fig.\ref{fig:pion_results}
our calculations for the pion involved processes.
Fig.\ref{fig:pion_results}(a) and Fig.\ref{fig:pion_results}(b) represent the resulting pion structure function and our results for the pion-nucleon and pion-pion total cross sections, respectively.
These results are obtained without any adjustable parameter, utilizing the bottom-up AdS/QCD model of mesons to calculate the pion density distribution.

\section{Summary}
In this report, we have presented our analysis on the DIS at small $x$ and the hadron-hadron scattering at high energies in the framework of holographic QCD.
Combining the BPST Pomeron exchange kernel and the density distributions obtained from the bottom-up AdS/QCD models of hadrons, we have calculated the structure functions and hadron-hadron total cross sections.
Our resulting nucleon structure functions and nucleon-nucleon total cross section agree with the experimental data, which implies that the present model setup can well describe the phenomena.
Since other applications of this framework are possible, further investigations are certainly needed.
Also, it is expected that our predictions presented here can be tested at the future experimental facilities.



\end{document}